\documentclass[usegraphicx,useAMS,usenatbib]{mn2e}

\newcommand{\beq}{\begin{equation}}
\newcommand{\beqa}{\begin{eqnarray}}
\newcommand{\eeq}{\end{equation}}
\newcommand{\eeqa}{\end{eqnarray}}

\newcommand{\bfk}{\mathbf{k}}

\newcommand{\bfp}{\mathbf{p}}
\newcommand{\bfq}{\mathbf{q}}

\title[Simulations of BAO I]{Simulations of Baryon Acoustic Oscillations I:
Growth of Large-Scale Density Fluctuations}
\author[R. Takahashi et al.]{Ryuichi Takahashi$^1$, Naoki Yoshida$^{1,2}$, Takahiko Matsubara$^1$, Naoshi Sugiyama$^{1,2}$, 
\newauthor Issha Kayo$^2$, Takahiro Nishimichi$^3$, Akihito Shirata$^{3,4}$,
 Atsushi Taruya$^{3,5}$,
\newauthor Shun Saito$^3$, Kazuhiro Yahata$^3$, and Yasushi Suto$^3$ \\
$^1$ Department of Physics and Astrophysics, Nagoya University, Chikusa,
 Nagoya 464-8602, Japan \\
$^2$ Institute for Physics and Mathematics of the Universe, University of
 Tokyo, 5-1-5 Kashiwa-no-ha, Kashiwa City, Chiba 277-8582, Japan \\
$^3$ Department of Physics, School of Science, The University of Tokyo,
 Tokyo 113-0033, Japan \\
$^4$ Department of Physics, Tokyo Institute of Technology, 
 Tokyo 152-8511, Japan \\
$^5$ Research Center for the Early Universe, The University of Tokyo, Tokyo 133-0033, Japan
}

\begin{document}

\date{}

\pagerange{\pageref{firstpage}--\pageref{lastpage}} \pubyear{2008}

\maketitle

\label{firstpage}

\begin{abstract}

We critically examine how well the evolution of large-scale density
perturbations is followed in cosmological $N$-body simulations. 
We first run a large volume simulation and perform a mode-by-mode 
analysis in three-dimensional Fourier space.
We show that the growth of large-scale fluctuations significantly
deviates from linear theory predictions. 
The deviations are caused by {\it nonlinear} coupling with a small 
number of modes at largest scales owing to finiteness of the simulation
volume. We then develop an analytic model based on second-order perturbation
theory to quantify the effect.  Our model accurately reproduces the
simulation results.  For a single realization, the second-order
effect appears typically as ``zig-zag'' patterns around the
linear-theory prediction, which imprints artificial ``oscillations''
that lie on the real baryon-acoustic oscillations.
Although an ensemble average of a number of
realizations approaches the linear theory prediction, the
dispersions of the realizations remain large even for a large
simulation volume of several hundred megaparsecs on a side.  For
the standard $\Lambda$CDM model, the deviations from linear growth
rate are as large as 10 percent for a simulation volume with $L = 500
h^{-1}$Mpc and for a bin width in wavenumber of $\Delta k = 0.005
h$Mpc$^{-1}$, which are comparable to the intrinsic variance of
 Gaussian random realizations. We find that the
 dispersions scales as $\propto L^{-3/2} \Delta k^{-1/2}$ and that 
 the mean dispersion amplitude can be made smaller than a percent only
 if we use a very large volume of $L > 2h^{-1}$Gpc.  The
 finite box size effect needs to be appropriately taken into
 account when interpreting results from large-scale structure
 simulations for future dark energy surveys using baryon acoustic 
 oscillations.
\end{abstract}


\section{Introduction}

Understanding the nature of dark energy that dominates the energy content of 
the universe is one of the main challenges in cosmology.
The time evolution of the mysterious dark component
is accessible only by astronomical observations.
Baryon acoustic oscillations (BAO) can be used as a standard ruler by which
precise measurement of the cosmological distance scale is achievable
(e.g., \citealt{eht98}; Seo \& Eisenstein 2003; Matsubara 2004).

Recent large galaxy redshift surveys, the Sloan
Digital Sky Survey and the 2-degree Field survey, detected the
signature of the baryon acoustic peaks and thus provide
 constraints on the dark energy
\citep{e05,c05,p07,o07}.  Future observational programs will utilize
the distribution of millions of high-redshift galaxies to detect BAO
with higher accuracy.  In order to properly interpret
  these observations, it is necessary to make accurate theoretical
predictions for the length scale and other characteristic features of
BAO (e.g. \citealt{n07,sss08}).  Theoretically, a crucial issue is the
non-linear evolution of matter and galaxy distributions (e.g.,
\citealt{se05,abfl07,gbs07,sss07}).  One usually resorts to using
cosmological $N$-body simulations for this, but various effects --both
physical and numerical-- need to be understood in order to extract
useful information.  First of all, the power spectrum
for a realization of a Gaussian random field has intrinsic deviations
from expected values at any wavenumber, i.e., the mode amplitudes are
Rayleigh-distributed (see e.g., \citealt{m07}). A realization may thus
show an additional oscillatory feature on large scales which
compromises the true BAO signature (Huff et al. 2007). There are also
a number of numerical issues.  Accurate time integration is necessary
in order to follow the evolution of large-scale density perturbations
which have small amplitudes.  Finite-box size limits the sampling of
wavenumbers at the largest scales, where the power
amplitude is dominated by only a few modes (\citealt{bp06}
 studied the finite box size effect on the mass function of dark matter
 halos.)

In this paper, we examine how accurately the evolution of large-scale
density perturbations is followed in standard cosmological $N$-body
simulations.  In particular, we study the
characteristic ``wiggle'' features which are often found in the matter
power spectra calculated from $N$-body simulations in previous
studies.  We use an approach based on perturbation theory to study
nonlinear effects in detail.
A further extensive study is presented in a separate paper by Nishimichi
et al. (in preparation).

Throughout the present paper, we adopt the standard $\Lambda$CDM model
with matter density $\Omega_{m} =0.241$, baryon density
$\Omega_{\rm b}=0.041$, cosmological constant $\Omega_{\Lambda}=0.759$,
spectral index $n_{\rm s}=0.958$, amplitude of fluctuations $\sigma_8=0.76$, 
and expansion rate at the present time $H_{0}=73.2$km s$^{-1}$ Mpc$^{-1}$,
consistent with the 3-year WMAP results (Spergel et al. 2007).

\section{Method}

\subsection{The cosmological simulations}
We use the cosmological simulation code Gadget-2 \citep{syw01,s05}.
For our fiducial runs, we employ $256^3$ particles in a volume of 
$L=500 h^{-1}$ Mpc on a side. 
We dump snapshots at a number of time steps (redshifts)
to study the evolution of the density power spectrum.
The simulation parameters are chosen such that sufficient convergence
is achieved in the measured power spectrum at the present epoch
(Takahashi et al., in preparation). 

We generate initial conditions for our runs based on the standard
Zel'dovich approximation using the matter transfer function calculated
by CAMB (Code for Anisotropies in the Microwave Background;
\citealt{lcl00}). The initial redshift is set to be $z_{\rm in}=30$.
When we generate a realization for a Gaussian random field, the
amplitude of each $k$-mode is assigned such that the ensemble follows
the Rayleigh distribution. While the mean of the
power is expected to approach the input value at $k$ for an ensemble
of large modes, the actual assigned power in a finite $k$-bin can
deviate significantly from the expected value.  Note also that a
Rayleigh distribution has a positive skew, which causes the median to
be smaller than the mean.

\subsection{Fourier mode analysis}
We first compute the density field for each output of the 
$N$-body simulation.  
We use the CIC (cloud-in-cell) interpolation when assigning particles
on grids. 
We check that the interpolation method does not affect
  the scales of interest ($k\lse 0.1$) by comparing various schemes.
We then apply a Fast Fourier Transform\footnote{FFTW Home page :
 http://www.fftw.org/} to obtain the density field $\delta(\bfk)$ in
three-dimensional Fourier space.
We will examine both the amplitudes and
the phases in detail in subsequent sections.

In order to study closely the Fourier mode-coupling, 
we calculate the mean amplitude of modes for a given realization
 with wavenumber vector
${\bf k} = (k_1,k_2,k_3)$ as
\begin{equation}
\hat{P}(k) = \frac{1}{N_k} \sum_{|\bfk|=k} \left| \delta ({\bf k}) \right|^2,
\label{pk}
\end{equation}
where the summation is for all the wavenumbers of
 $|\bfk|=k=(k_1^2+k_2^2+k_3^2)^{1/2}$,
 $N_k$ is the number of modes in $k$, and the wavenumber is discretized
 as $k_i = (2 \pi/L) n_i$ with an integer $n_i$. 
An ensemble average of a number of realizations provides its
 expectation value of $P(k)=\langle \hat{P}(k) \rangle$.

In order to study the {\it evolution} of power spectrum,
we divide the measured power spectrum in equation
 (\ref{pk}) at redshift $z$ by the initial one at $z_{\rm in}=30$,
 and then multiply it by the input power spectrum.
In this way, the initial random scatter included in the power spectrum
is removed.

\section{Results}

\begin{figure}
  \includegraphics[width=80mm]{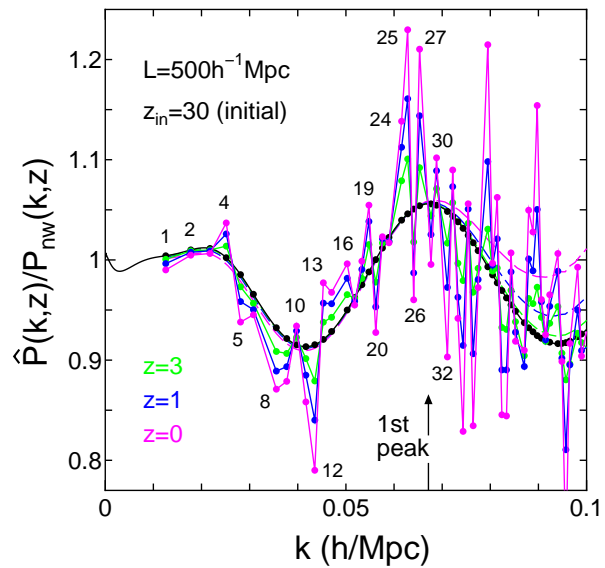}
  \caption{We plot the evolution of the power spectrum from the initial 
epoch (black line) to $z=3$ (green), $z=1$ (blue), and $z=0$ (purple). 
The measured power spectrum is divided by
the no-wiggle model of Eisenstein \& Hu (1998).
We subtract the intrinsic deviations from the input power spectrum at
 the initial epoch. 
The numbers indicate integer sums of $n_1^2+n_2^2+n_3^2$ of wavenumber
 vectors. 
The dashed lines are the one-loop power spectra at each redshift
(see text).
 }
  \label{CAMB}
\end{figure}

Fig.\ref{CAMB} shows the evolution of power spectrum ${\hat{P}} (k)$
 for a single realization.
We show the mean amplitude for modes which have exactly the same 
wavevector norm, $|\bfk|^2 = k_{1}^{2} + k_{2}^{2} + k_{3}^{2}$,
rather than binning in $k$. 
The vertical axis is the power spectrum divided by the no-wiggle model
of Eisenstein \& Hu (1999). 
The black line with symbols is the linear theory prediction
 with CAMB.
The green, blue, and purple lines with dots 
are the measured mean values at each wave number
at $z=3,1,$ and $0$, respectively.
The numbers in the figure indicate
integer sums of $n_1^2+n_2^2+n_3^2$ of wavenumber vectors. 

As clearly seen in the figure, the power amplitudes 
deviate from the linear theory prediction at low redshifts.
The deviations appear to grow in time monotonically.
Some modes (e.g. $n^2=4,13,19,25,27$) grow more rapidly than 
 the linear growth, while other modes (e.g. $n^2=8,12,20,26,32$)
 grow less. 
These features can be seen even in higher resolution simulation of
 \cite{s05b} (see their Fig.6).
Since the initial randomness of the amplitude of each mode
has been already subtracted in the figure as described in section 2.2, 
the remaining differences 
plotted in Fig. 1 are due either to numerical integration errors
or to some unknown physical effects.
The deviations are indeed large, with the amplitudes
being more than $10\%$ at the scale of the first-peak of the BAO.
It is thus important to understand and correct the apparent oscillatory
 features if these are artificial effects. 

In the next section, we show that the deviations are {\it not} 
owing to numerical integration errors {\it but} due to the finite number
of modes at the largest scales. 
We use second-order perturbation theory to explain 
the systematic deviations.

\section{Perturbation theory}

Second-order perturbation theory describes the evolution
of a density perturbation as (e.g. \citealt{bcgs02})
\beq
  \delta(\bfk,z)=\frac{D(z)}{D_{\rm in}} \delta_1(\bfk) +
  \left( \frac{D(z)}{D_{\rm in}} \right)^2 \delta_2(\bfk),
\label{delta}
\eeq
where $\delta_1(\bfk)$ and $D_{\rm in}$ are the linear density
and the linear growth factor evaluated at the initial redshift.
The second-order term is given by
\beq
    \delta_{2}(\bfk)= \sum_\bfp F_2(\bfp,\bfk-\bfp) \delta_{1}(\bfp)
  \delta_{1}(\bfk-\bfp),
\label{mc}
\eeq
with
\beq
   F_2(\bfp,\bfq)=\frac{5}{7}+\frac{\bfp \cdot \bfq}{2} \left( \frac{1}{p^2}
 + \frac{1}{q^2} \right) + \frac{2}{7} \frac{\left( \bfp \cdot \bfq \right)^2}
   {p^2 q^2}.
\label{f2}
\eeq We sum up all the modes up to the Nyquist frequency
($256^3$ modes in total) in equation
(\ref{mc}). 
Here, equation (\ref{f2}) includes the fastest growing mode.
\cite{bcs08} recently present the correct formula of $F_2$ including
 the sub-leading growing mode.

Let us explicitly write the amplitude and the phase
of a mode as
\beq
\delta(\bfk,z)=\left| \delta(\bfk,z) \right| {\mbox{exp}} \left(
 {\mbox{i}} \phi(\bfk,z) \right),
\eeq
Then the evolution of amplitude in each mode is
\beqa
 \frac{\hat{P}(k,z)/\hat{P}(k,z_{\rm in})}{D(z)^2/D_{\rm in}^2}
 = 1+ \frac{1}{N_k} \sum_{|\bfk|=k} 2 \mbox{Re} \left[ \delta_1(\bfk)
 \delta_2^*(\bfk) \right] \nonumber \\
 \times  \frac{1}{\hat{P}(k,z_{\rm in})}
 \frac{D(z)}{D_{\rm in}},
\label{amp_evolv}
\eeqa
whereas the phase evolution is  
\beqa
  \phi(\bfk,z) - \phi_{\rm in}(\bfk) = \sin \phi_{\rm in}(\bfk)
  \cos \phi_{\rm in}(\bfk) \nonumber \\
 \times \left( \frac{\mbox{Im}\delta_2(\bfk)}
 {\mbox{Im}\delta_1(\bfk)} - \frac{\mbox{Re}\delta_2(\bfk)}
 {\mbox{Re}\delta_1(\bfk)} \right) \frac{D(z)}{D_{\rm in}},
\label{phase_evolv}
\eeqa
up to second order.
The expressions in equations (\ref{amp_evolv}) and (\ref{phase_evolv})
 are independent of the initial redshift for the late time
 ($D \gg D_{\rm in}$), since $\delta_1 \propto D_{\rm in}$ and
 $\delta_2 \propto P(k,z_{\rm in}) \propto D_{\rm in}^2$.
We do not distinguish between $\delta$ and $\delta_1$ at the initial redshift
 ($z_{\rm in}=30$), since $\delta_2$ is much smaller than
 $\delta_1$ at that time.
\footnote{Nishimichi et al. (in preparation) distinguish $\delta$ from
 $\delta_1$ at the initial epoch with the 2LPT initial condition
 (\citealt{cps06})} and provide more detail analysis.

It is clear from equation (\ref{f2}) that nonlinear mode-coupling occurs with
 particular sets of wavenumber vectors such that
$\bfp + \bfq = \bfk$. From equation (\ref{f2}), we obtain
\beq
  F_2(\bfp,\bfk-\bfp) \rightarrow \left( \frac{3}{14} - \frac{5}{7}
 \cos^2 \theta \right) \frac{k^2}{p^2},
\label{f2_1}
\eeq
for $k \ll p$, and 
\beq
 F_2(\bfp,\bfk-\bfp) \rightarrow \frac{1}{2} \frac{k}{p} \cos \theta,
\eeq
for $k \gg p$. Here $\theta$ is an angle between $\bfk$ and $\bfp$.
Hence the coupling to the mode of much smaller scale $p (\gg k)$ is
 negligibly weak, while the coupling to much larger scale $p (\ll k)$
 is strong. 
In summary, most of the contribution 
to the second-order evolution of a mode 
comes from the modes of comparable scales or larger.\footnote{
\citealt{m88} examined the growth of the small-scale perturbation on
 the background of the large-scale perturbation.}

For a Gaussian random field, the mode amplitudes are
Rayleigh-distributed, and thus there is a finite probability that a
mode has a very large or a very small amplitude with respect to the
expected {\it mean} value.  Some peculiar modes, which have very large
or very small amplitudes compared to the mean, strongly affect the
 growth of other modes through the
mode-coupling 
as described in the above.

\begin{figure*}
  \includegraphics[width=66mm]{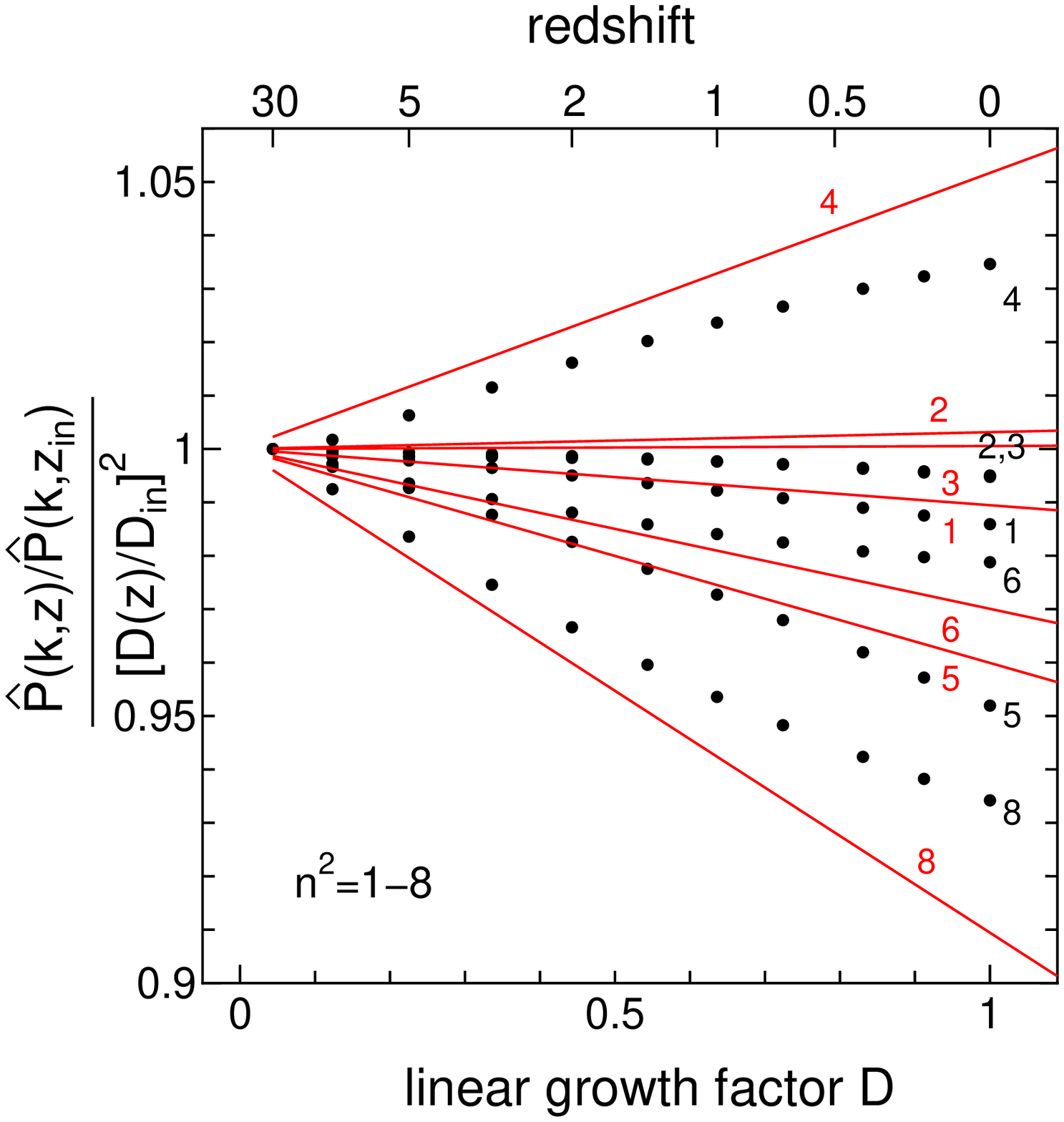}
  \hspace{0.5cm}
  \includegraphics[width=66mm]{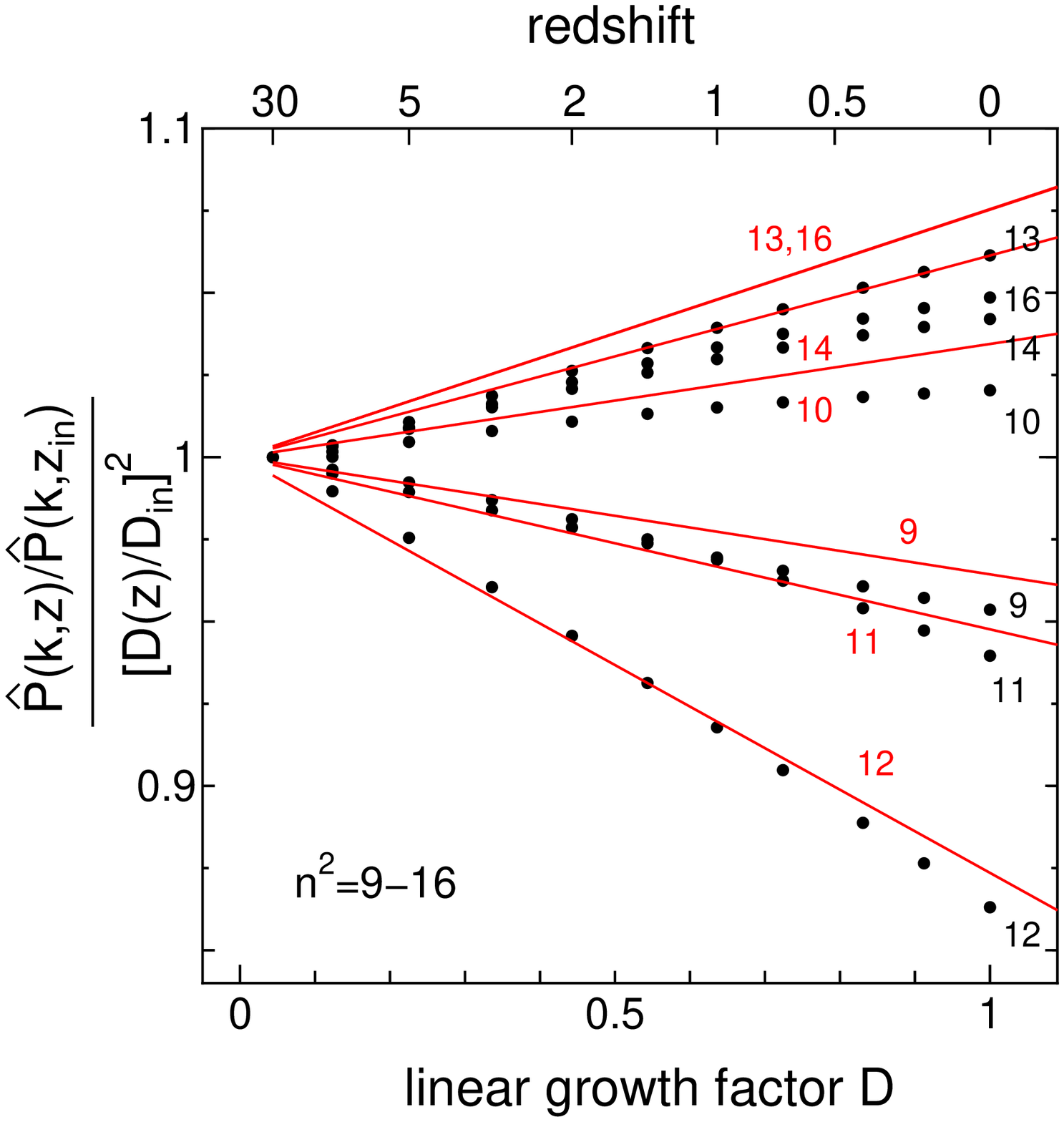}
  \includegraphics[width=66mm]{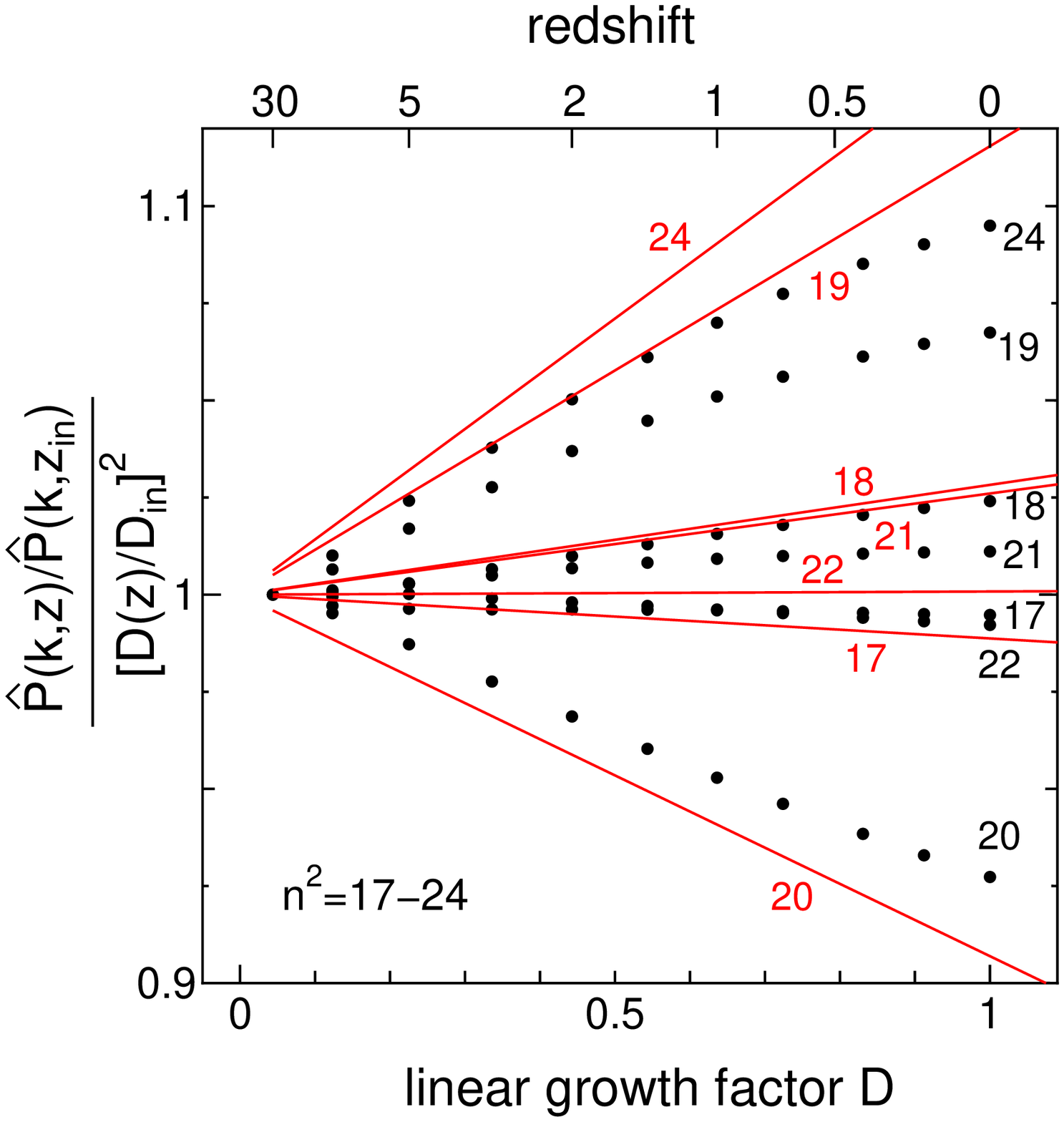}
  \hspace{0.5cm}
  \includegraphics[width=66mm]{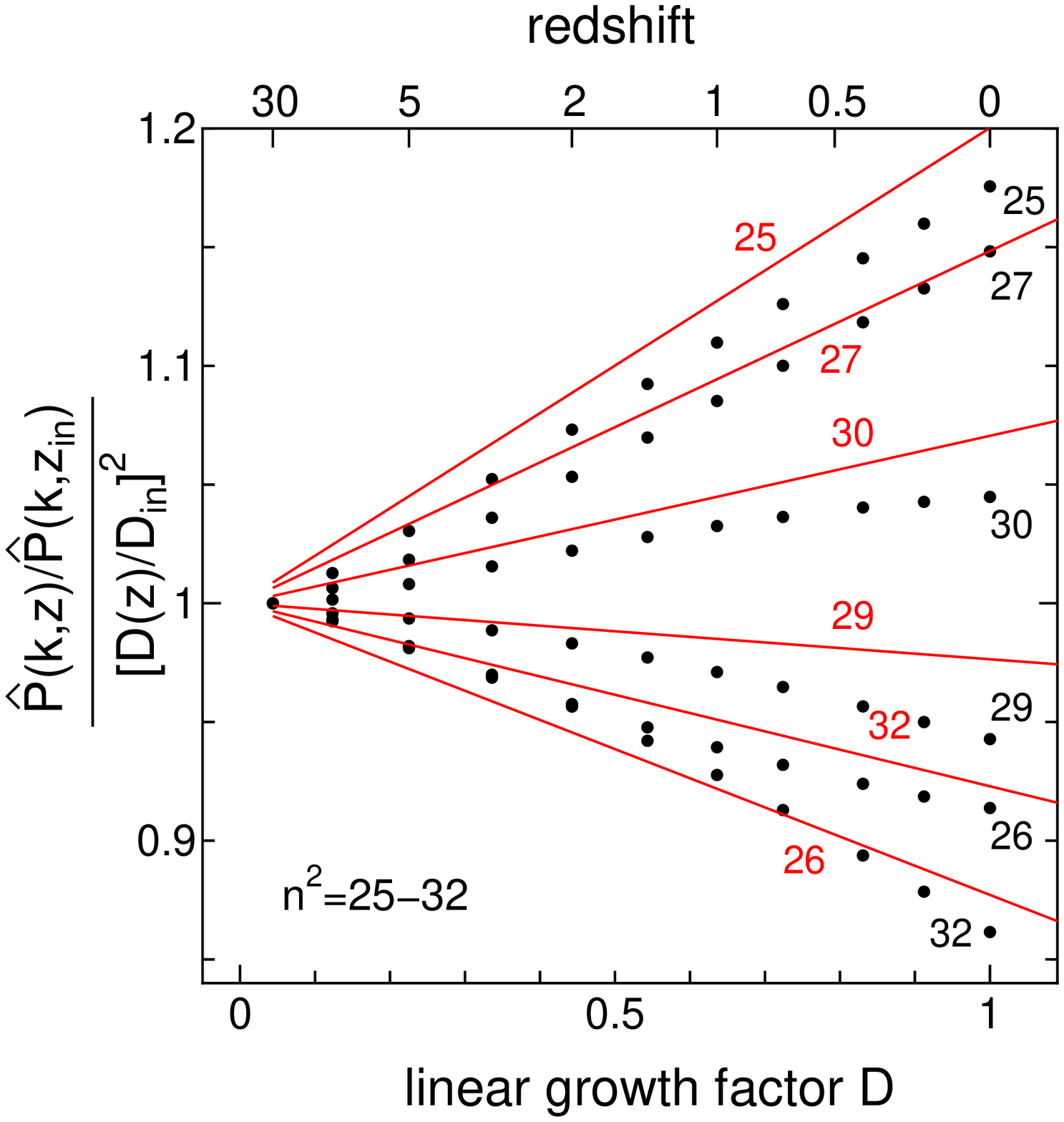}
  \caption{Evolution of the deviation of the power amplitude 
  with respect to the linear theory prediction.
  The dots are the measurements from our simulation, 
  and red solid lines are the model prediction using the 
  second-order perturbation theory.
 The integers denote $n^2=n_1^2+n_2^2+n_3^3$ of wavenumbers, and
the figures show different range of $n^2$,
  $n^2=1-8$ (upper left panel), $n^2=9-16$ (upper right panel), 
  $n^2=17-24$ (lower left panel), and $n^2=25-32$ (lower right panel).}
  \label{mode_evolv}
\end{figure*}

\begin{figure*}
  \includegraphics[width=66mm]{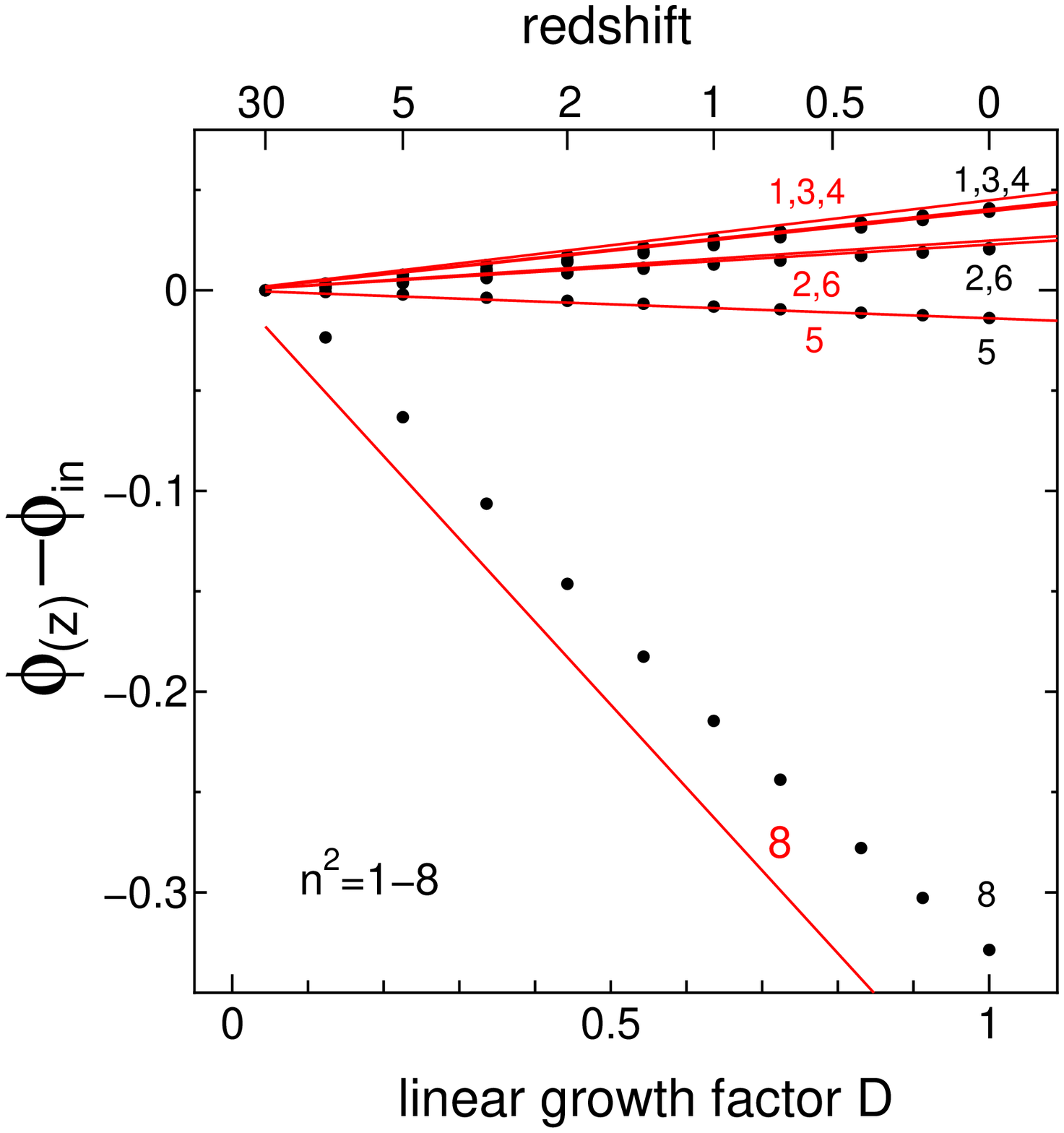}
  \hspace{0.5cm}
  \includegraphics[width=66mm]{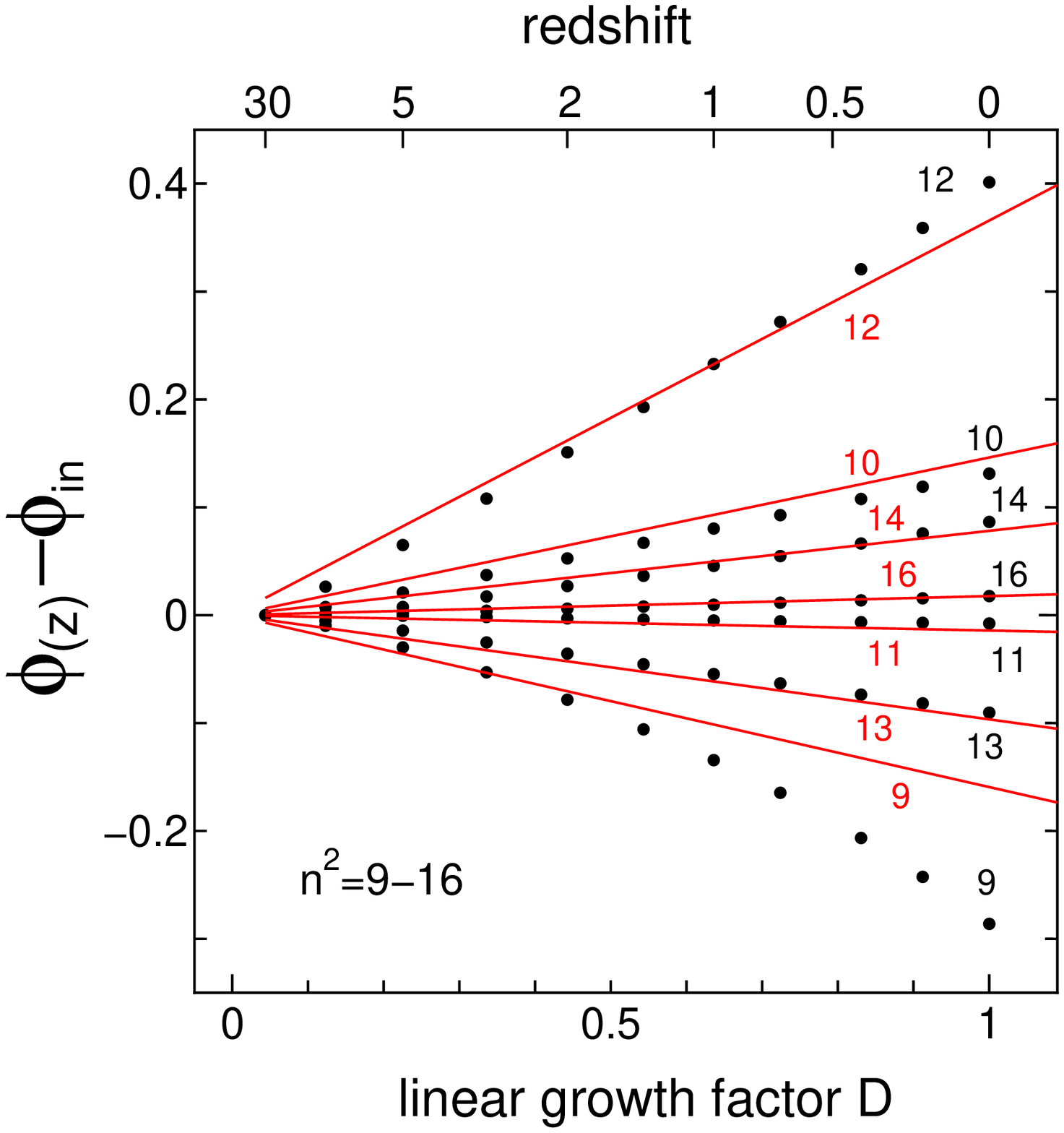}
  \includegraphics[width=66mm]{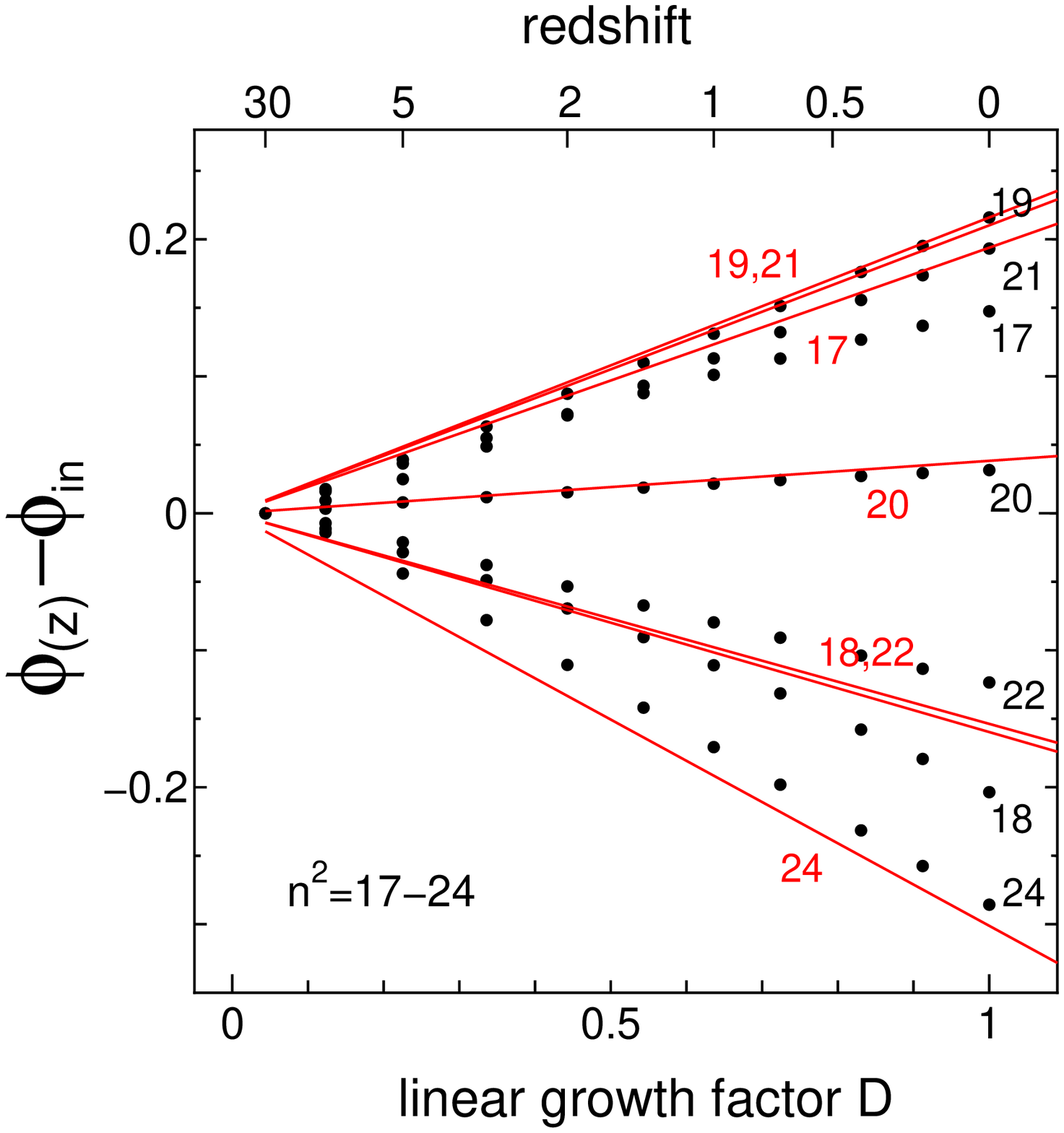}
  \hspace{0.5cm}
  \includegraphics[width=66mm]{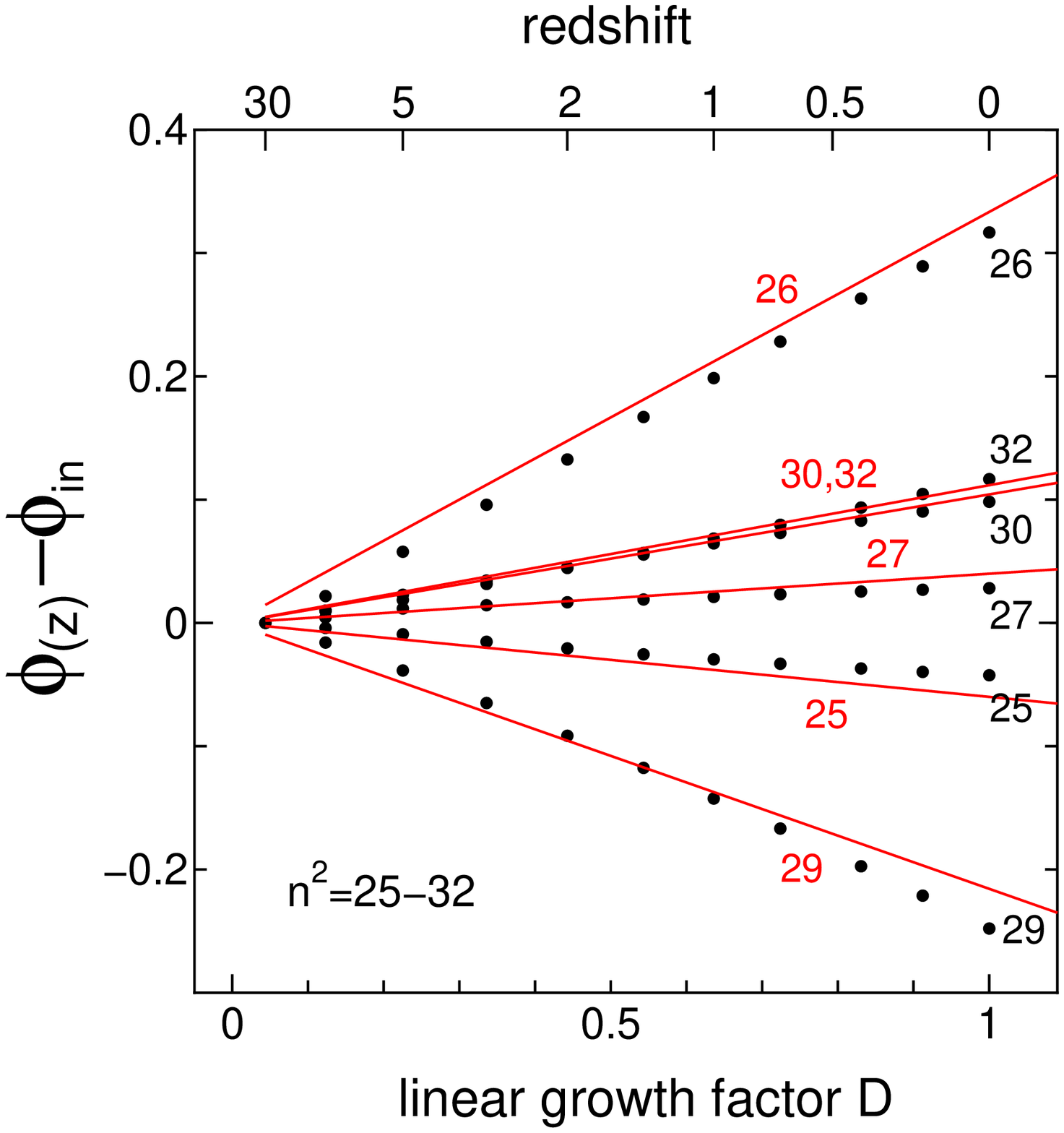}
  \caption{Same as Fig.\ref{mode_evolv}, but for phase evolution
    in units of radians.
  We plot the results only for modes with $n_1 \geq n_2 \geq n_3$.
}
  \label{mode_evolv2}
\end{figure*}

In an ideal situation where there are infinite number of modes, the
second term in equation (\ref{amp_evolv}) vanishes.  In that case, the
leading correction arises from the forth order of $\delta_1$.  Then
the resultant power spectrum with the one-loop
correction is, 
\beq P_{\rm 1loop}(k,z)=\left( \frac{D(z)}{D_{\rm in}} \right)^2
P_{11}(k) + \left( \frac{D(z)}{D_{\rm in}} \right)^4 \left[ P_{22}(k)
    + P_{13}(k) \right],
\label{one-loop}
\eeq
where $P_{11}=\left< | \delta_1 |^2 \right>$,
  $P_{22}=\left< | \delta_2 |^2 \right>$, $P_{13}= 2 \left< {\rm Re} [
      \delta_1 \delta_3^* ] \right>$ \citep{mss92,jb94,jk06}.  We
  integrate from $k=2\pi/L$ to the Nyquist frequency in the
  calculation of $P_{22}$ and $P_{13}$.

The dashed lines in Fig.\ref{CAMB} are the one-loop power spectrum
at each redshift.
It suggests that the linear theory is applicable for $k < 0.07 h$/Mpc at $z=0$.
However, the finite mode coupling in the second term of
 equation (\ref{amp_evolv}) significantly
 changes the evolution of the power spectrum even in the linear
 regime.\footnote{\cite{s99} also investigated the finite mode effect on
 the one-loop correction terms, $P_{22}+P_{13}$, in equation
 (\ref{one-loop}).}

Fig.\ref{mode_evolv} shows the evolution of
 the mean amplitude of modes with identical wavenumber
  $n^2$ in the range of $1-32$.  Here, $n^2 \simeq 30$ corresponds to
the position of the first peak (see Fig.\ref{CAMB}).  The four panels
are for $n^2=1-8$ (upper left panel), $n^2=9-16$ (upper right panel),
$n^2=17-24$ (lower left panel), and $n^2=25-32$ (lower right panel).
The dots are the measurement from simulation outputs, and red solid
lines are the theoretical prediction from the initial density fields
at $z_{\rm in}=30$ in equation (\ref{amp_evolv}).  The second-order
perturbation theory reproduces the simulation results rather well.
The theory fits the data within $0.5 \%$ at $z=2$ and $2 \%$ at $z=0$
for larger scale ($n^2=1-8$), whereas within $1 \%$ at $z=2$ and $10
\%$ at $z=0$ for smaller scale ($n^2=25-32$).  This is because the
second order perturbation theory is applicable at large scales and/or
at high redshift.

Fig.\ref{mode_evolv2} is the same as Fig.\ref{mode_evolv}, but for phase
 evolution.
We plot the results only for modes with $n_1 \geq n_2 \geq n_3$,
because the mean of the phase at $k$, $\sum \phi(\bfk)$,
 is zero (since $\phi(\bfk) + \phi(-\bfk)=0$).
The phase shifts are typically $\approx 0.1$ radian at $z=0$.
Perturbation theory well reproduces the results.
Even if there are infinite modes, the right hand side of 
 equation (\ref{phase_evolv}) still remains.
The phase shift is not due to the finite box size effect.

Previously \cite{rg91} and \cite{g92} studied the evolution of
amplitude and phase in each mode using two
dimensional simulations.  They also calculated
second-order perturbation theory and found the
deviation from the linear theory grows in proportional to the scale
factor in the EdS model.  \cite{ss91}, \cite{ss92}
and \cite{jb98} also examined the nonlinear evolution in each mode.
However they did not compare the theoretical prediction with the
simulation results in detail.  Their motivations were to understand
the evolution of the density fluctuations in the
nonlinear regime, whereas our interest here is in the growth of
perturbations at the {\it linear} scale.

\section{Statistical Analysis}
The previous section considers second-order effects
for a single realization.  In this section we run $100$ simulations
 to calculate dispersions of amplitude and phase deviations from
linear theory.  We prepare the $100$ realizations
for each of three box sizes of $L = 500 h^{-1}$Mpc, $1
  h^{-1}$Gpc, and $2 h^{-1}$Gpc, and $z_{\rm in}=30,20$ and $10$,
  respectively.

\begin{figure}
  \includegraphics[width=84mm]{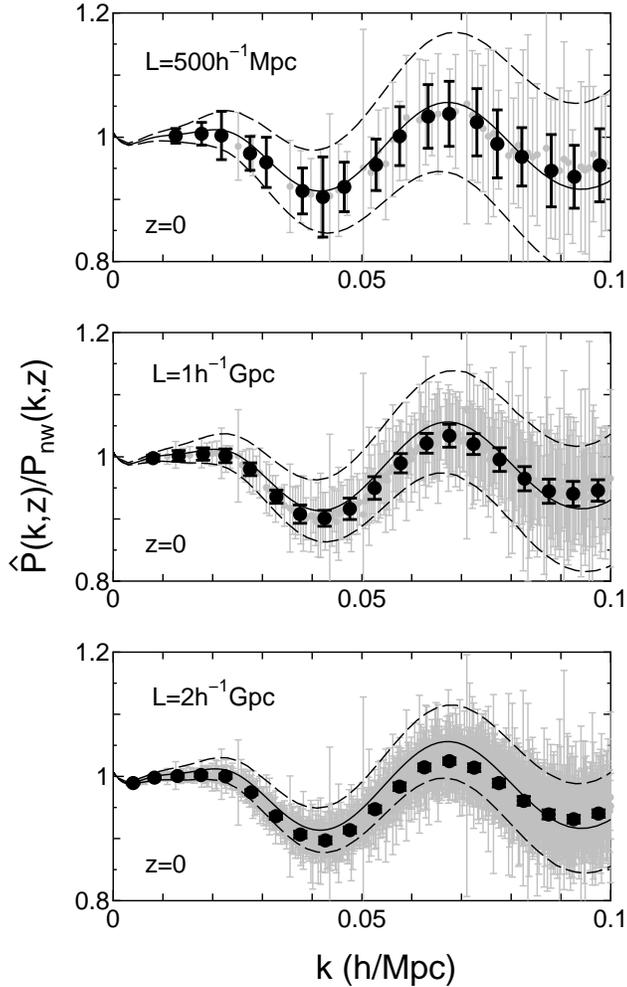}
  \caption{The amplitude dispersions of the $100$ realizations at $z=0$
 for $L=500 h^{-1}$ Mpc (top), $1 h^{-1}$ Gpc (middle), and $2 h^{-1}$
 Gpc (bottom).
 The grey dots with error bars are for the un-binned data, while the
 black big symbols are for the binned data of $\Delta k=0.005 h$/Mpc.
 The value of $k$ for the binned data is the weighted mean of $k$ with
 the number of wavenumbers in the bin.
 The dashed lines are the theoretical prediction.}
  \label{pk_disp}
\end{figure}

Fig.\ref{pk_disp} shows the remaining amplitude
  dispersions from the linear theory prediction after correcting for
  the initial randomness at $z=0$ for $L=500 h^{-1}$ Mpc (top), $1
h^{-1}$ Gpc (middle), and $2 h^{-1}$ Gpc (bottom).  Since we already
subtract the initial deviations due to the Gaussian distribution, the
residuals arise from the mode-coupling during the evolution.  The grey
dots with error bars are the means with $1 \sigma$ scatters.  By using
a sufficiently large number of realizations, the means converge to the
true values (solid line), and the magnitude of the dispersions is
insensitive to the number of realizations.  For $L=500 h^{-1}$Mpc, the
dispersions are $\sim 10 \%$ near the first peak, and $\sim 5 \%$ even
for a very large volume of $2 h^{-1}$ Gpc on a side.  The dashed lines
show the theoretical prediction of the $1 \sigma$ scatter, which is
the rms (root-mean-square) of the second term in equation
(\ref{amp_evolv}) :
\beqa
 \sigma_{\rm amp}^2 &\equiv& \left< \left(  \frac{\hat{P}(k,z)
 /\hat{P}(k,z_{\rm in})}{D(z)^2/D_{\rm in}^2} -1 \right)^2 \right> 
 \nonumber \\
 &=& \frac{4 P_{22} (k,z_{\rm in})}{P_{11}(k,z_{\rm in})}
 \frac{1}{\Delta N_k} \left( \frac{D(z)}{D_{\rm in}} \right)^2.
\label{sigma_amp}
\eeqa 
Here, $\Delta N_k$ is the number of modes in the bin, $\Delta
N_k=4 \pi n^2 \Delta n$ with $n=(L/2 \pi) k$.  In this unbinning case,
the number of modes is $\Delta N_k = k L \Delta n^2$ (with $\Delta
n^2=1$).  The dashed lines well reproduce the results.

Fig.\ref{pk_disp} also shows the results for the binned data of
 $\Delta k=0.005 h$/Mpc by the black big symbols.
In this case, we use the power spectrum defined as
 $\hat{P}(k)=(1/\Delta N_k) \sum \left| \delta(\bfk) \right|^2$, summing up
 all the modes between $(k-\Delta k/2,k+\Delta k/2)$, instead of
 equation (\ref{pk}).
Here, the number of modes in the bin is
\beq
  \Delta N_k=(L^3 k^2)/(2 \pi^2) \Delta k.
\label{nk_bin}
\eeq
We calculate the means and error bars for the binned $\hat{P}(k)$.

\begin{figure}
  \includegraphics[width=80mm]{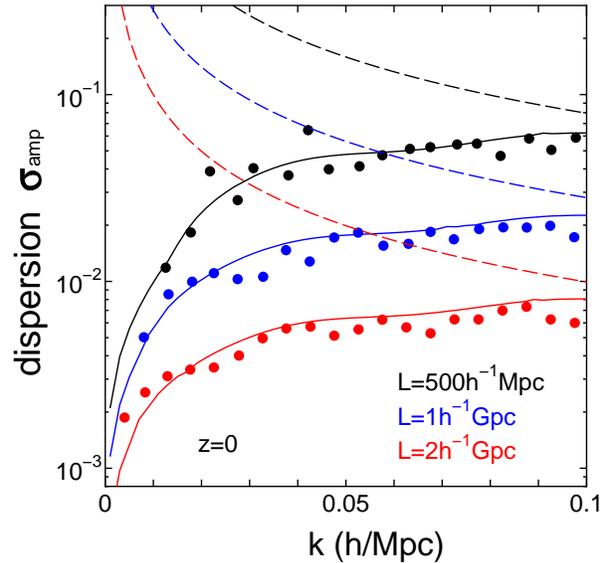}
  \caption{The amplitude dispersions calculated from our simulation outputs
 (filled circle $\bullet$)
 and the theoretical predictions (solid lines).
 We also show the dispersions due to the initial Gaussian distribution
 (dashed lines). 
 The vertical dotted line is the position of the BAO first peak.
}
  \label{pk_disp_bin}
\end{figure}

Fig.\ref{pk_disp_bin} shows the amplitude dispersions calculated from our
 simulation outputs for $\Delta k=0.005 h$/Mpc
 (filled circle) and the theoretical prediction (solid line).
From this figure with equations (\ref{sigma_amp}) and (\ref{nk_bin}),
 we find that the dispersion is approximated as
\beq
  \sigma_{\rm amp} (z=0) \simeq 2 ~\% \left( \frac{L}{1 {\mbox{Gpc/h}}}
 \right)^{-3/2}
 \left( \frac{\Delta k}{0.005 {\mbox{h/Mpc}}} \right)^{-1/2},
\label{sigma_amp2}
\eeq at $k=0.02-0.1h$/Mpc.  The dispersion is proportional to $\Delta
N_k^{-1/2} \propto L^{-3/2} \Delta k^{-1/2}$ from equation
(\ref{nk_bin}).  Note that even with a large simulation volume of $L
\sim 1$ Gpc with $k$-binning, the dispersions still
  remain at the level of a few percent.

\begin{figure}
  \includegraphics[width=80mm]{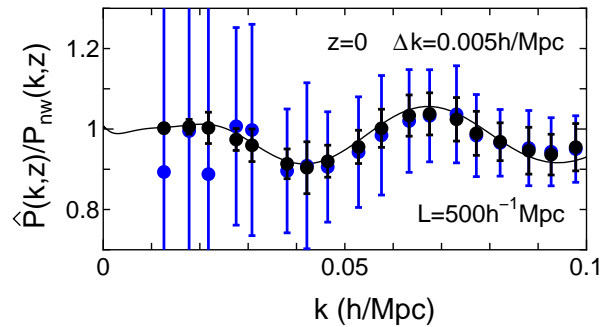}
  \caption{We compare two dispersions.
Blue points with error bars show intrinsic scatter around the
 expected mean power spectrum for initial Gaussian random density fields.
Black points show the dispersions owing to the finite nonlinear
 mode-coupling effect. 
}
  \label{pk_disp_gauran}
\end{figure}

So far we have discussed the amplitude of deviations from linear
theory.  Here we also consider the intrinsic scatter of the initial
Gaussian random realizations. In Fig.\ref{pk_disp_bin} the dashed line
is the dispersion for the initial distribution, which is given by
\footnote{The number of modes $\Delta N_k$ is divided by $2$ because
  the Fourier modes of $\delta(\bfk)$ and $\delta(-\bfk)$ are not
  independent.}  $(\Delta N_k/2)^{-1/2}$.  Fig.\ref{pk_disp_bin} shows
that the dashed lines decrease as $\propto (\Delta N_k)^{-1/2} \propto
k^{-1}$, while the solid lines increase because $P_{22}/P_{11}$
increases (see equation [\ref{sigma_amp}]).  These two dispersions are
comparable at $k \simeq 0.1 h$/Mpc where $2 P_{22}/P_{11} \simeq 1$ at
$z=0$.
 About a half of the dispersions near the position of the BAO first
 peak ($k \sim 0.07h$/Mpc) are attributed to the second-order effects.
 The result suggests that, at large scales, $k < 0.1h$/Mpc, the
 dispersions arise mainly from the initial Gaussian random
 distribution, while at smaller scale $k > 0.1h$/Mpc they are from the
 mode-coupling (based on the second or higher order perturbation)
 during the evolution.  In Fig.\ref{pk_disp_gauran}
   the blue symbols are the results for our 100 realizations.  The
   black symbols are same as in the top panel of Fig.\ref{pk_disp} for
   $\Delta k=0.005h$/Mpc.  As expected, the initial random
   realizations have larger scatters around the mean expected power
   spectrum, especially at the largest scales.

\begin{figure}
  \includegraphics[width=80mm]{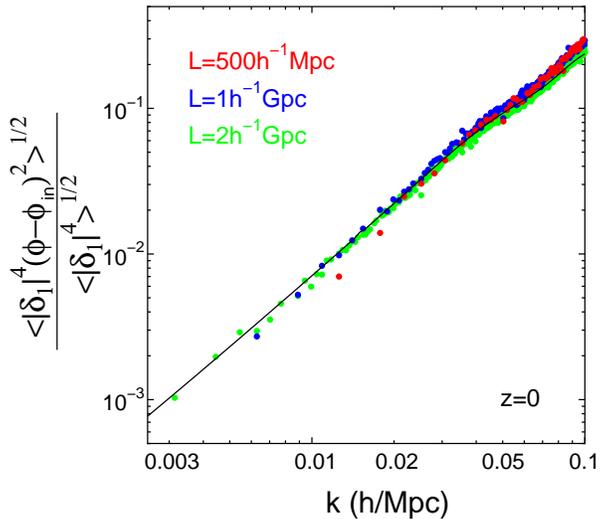}
  \caption{The phase dispersion of the $100$ realizations.
  The solid line is the theoretical prediction.}
  \label{th_disp}
\end{figure}

We have also performed a similar analysis
for the evolution of the mode phases (equation [7]).
Fig.\ref{th_disp} shows the phase dispersion calculated from
 our simulations (the dots).
Here we set
 $- \pi \leq \left( \phi - \phi_{\rm in} \right) \leq \pi$
 and calculate $\langle |\delta_1|^4 (\phi-\phi_{\rm in})^2 \rangle$
 instead of $\langle (\phi-\phi_{\rm in})^2 \rangle$.
This is because $(\phi-\phi_1) \propto 1/\delta_1$ in Eq.(\ref{phase_evolv})
 and its dispersion diverges at $\delta_1=0$.
We obtain the phase dispersion from equation (\ref{phase_evolv})
 as,\footnote{\cite{jb96} previously derived equation (\ref{sigma_th}) 
 with an approximation for the long-wave mode coupling.}
\beq
 \frac{\left< \left| \delta_1(\bfk) \right|^4 \left[ \phi(\bfk,z)-
 \phi_{\rm in}(\bfk) \right]^2 \right>}{\left< \left| \delta_1(\bfk)
 \right|^4 \right>}
 = \frac{P_{22} (k,z_{\rm in})}{6 P_{11}(k,z_{\rm in})}
  \left( \frac{D(z)}{D_{\rm in}} \right)^2.
\label{sigma_th}
\eeq
The solid lines are the theoretical prediction, which fit the simulation
 results well.
The phase dispersion in equation (\ref{sigma_th}), as well as
 the amplitude dispersion in equation (\ref{sigma_amp}), are
 independent of the initial redshift.
In the non-linear limit of $k \rightarrow \infty$, the phases are
 distributed randomly, and the phase dispersion
 approaches to $\pi/\sqrt{3}$ rad (e.g. \citealt{rg91}).

\section{Discussion and conclusions}

In this paper, we critically examined how accurately
  cosmological $N$-body simulations describe the evolution of
  large-scale density distributions, particularly focusing on the
  linear and/or quasi-linear scales. For the power spectrum calculated
  from a single realization, we found that the growth of large-scale
  fluctuations significantly deviates from the linear theory
  prediction, and the enhanced or suppressed growth of perturbations
  produces an ugly noisy pattern in the matter power spectrum.  This
  deviation is not due to the numerical errors in the $N$-body code,
  but due to the non-linear coupling between finite numbers of modes
  originating from the finite box size. To study the effect of the
  finite mode-coupling in detail, we developed perturbation theory and
  quantitatively estimated the finite-mode coupling to the power
  spectrum amplitude.  Mode-by-mode analysis in three-dimensional
  Fourier space reveals that the finite mode-coupling from the
  second-order perturbation is sufficient to explain the deviation
  from linear theory prediction on large scales. The dispersion of the
  mode-coupling effects estimated from second-order perturbation
  scales as $\propto L^{-3/2}\Delta k^{-1/2}$, and this may surpass
  the intrinsic scatter of the initial Gaussian distribution. Since the
  finite mode-coupling does not vanish even for a large-volume
  simulation, it is of critical importance to correct it properly for
  high-precision studies of baryon acoustic oscillations. 

\begin{figure}
  \includegraphics[width=80mm]{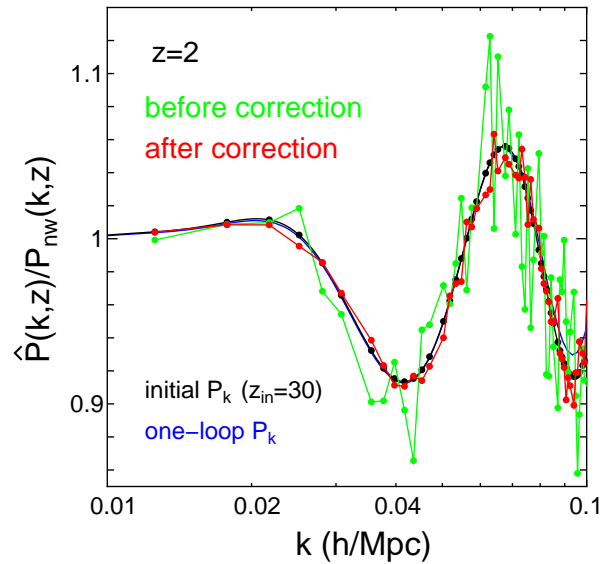}
  \caption{The power spectrum at $z=2$.
 The green line is the simulation output. 
 In the red line, we subtract the second-order
 perturbation contribution from the simulation output.
 The blue line is the one-loop power spectrum.
} 
\label{CAMB_corrected}
\end{figure}

We show that the perturbative approach is very helpful to quantify 
the significance of finite-mode coupling and this can be utilized as 
an efficient and powerful tool to correct the finite-mode coupling. 
As an example, in Fig.~\ref{CAMB_corrected}, we evaluate the 
power spectrum directly obtained from a single realization at 
$z=2$, and subtract the finite-mode coupling using the second-order 
perturbation. Compared 
the result before subtraction with that after subtraction, 
the deviation from linear theory is dramatically reduced and the 
noisy structures are effectively wiped out. As a result, 
even the single realization data of $N$-body simulation 
faithfully reproduces the linear theory prediction on large scales.

Although the present paper mainly concerns with the
  second-order perturbation theory, higher-order perturbations are
  also important for the relevant scales of the measurement of baryon
  acoustic oscillations, where the acoustic signature tends to be
  erased by the effect of non-linear clustering (e.g.
  \citealt{mc07,cs07,m07b,th08}). 
  The height of the first peak is found to be reduced about $2 \%$
  (J. Wang, A. Szalay et al. in preparation). Thus, the inclusion of the
  higher-order terms may be important for the estimation of the
  finite-mode coupling, which would be helpful to further reduce the
  noisy structures on small scales.

We note that the variance of the growth of
 matter power spectrum with respect to the linear theory
 prediction, $\langle [(\hat{P}/\hat{P}_{\rm in})/(D/D_{\rm in})^2-1]^2
 \rangle$, which we have studied,
 is different from the variance of the power spectrum itself,
 $\langle (\hat{P}-P)^2 \rangle$. It remains unclear if the numerical
 effects studied here are important in evaluating covariance matrices
 (e.g., \citealt{szh99,mw99,ns07})
 In future work, we will study nonlinear and numerical effects in the
 power spectrum covariance using a large set of simulations
 and analytic models.

\section*{Acknowledgments}
We thank Jie Wang, Erik Reese, and Simon White for useful comments
 and discussions.
We also thank the anonymous referee for careful reading and
 useful suggestions. 
This work is supported in part by Grant-in-Aid for Scientific Research
on Priority Areas No. 467 ``Probing the Dark Energy through an
Extremely Wide and Deep Survey with Subaru Telescope'', by the
Mitsubishi Foundation,
and by Japan Society for Promotion of Science (JSPS) Core-to-Core
Program ``International Research Network for Dark Energy'', and by
Grant-in-Aids for Scientific Research
(Nos.~18740132,~18540277,~18654047).  T. N., A. S. and K. Y. are
supported by Grants-in-Aid for Japan Society for the Promotion of
Science Fellows.

\bsp

\label{lastpage}

\end{document}